\documentclass[epj]{svjour}
\usepackage{graphics}
\usepackage{graphicx}
\usepackage{amsmath}
\usepackage{amssymb}

\begin{document}

\title{Magnetic properties of spin-$1/2$ Fermi gases with ferromagnetic interaction}

\author{Baobao Wang\inst{1} \and Jihong Qin\inst{1,} \thanks{\emph{jhqin@sas.ustb.edu.cn} }
\and Huaiming Guo\inst{2}}

\institute{Department of Physics, University of Science and
Technology Beijing, Beijing 100083, P.R. China \and Department of
Physics, Beihang University, Beijing 100191, P.R. China}

\date{Received: date / Revised version: date}

\abstract {We investigate the magnetic properties of spin-$1/2$
charged Fermi gases with ferromagnetic coupling via mean-field
theory, and find the interplay among the paramagnetism, diamagnetism
and ferromagnetism. Paramagnetism and diamagnetism compete with each
other. When increasing the ferromagnetic coupling the spontaneous
magnetization occurs in a weak magnetic field. The critical
ferromagnetic coupling constant of the paramagnetic phase to
ferromagnetic phase transition increases linearly with the
temperature. Both the paramagnetism and diamagnetism increase when
the magnetic field increases. It reveals the magnetization density
$\bar M$ increases firstly as the temperature increases, and then
reaches a maximum. Finally the magnetization density $\bar M$
decreases smoothly in the high temperature region. The domed shape
of the magnetization density $\bar M$ variation is different from
the behavior of Bose gas with ferromagnetic coupling. We also find
the curve of susceptibility follows the Curie-Weiss law, and for a
given temperature the susceptibility is directly proportional to the
Land\'{e} factor.}

\maketitle

\section{Introduction}
Magnetism of electron gases has been one of the central issues in
condensed matter physics. At low temperature, Fermi particles fill
the Fermi level from the lowest energy values, subject to the Pauli
exclusion principle, which is different from the Bose gas. Since the
observation of Bose-Einstein condensation, the trapped ultracold
Fermi gases have attracted great interest
\cite{Noronha,Giorgini,Bloch,Greif}. The measurement of magnetic
susceptibility for ultracold Fermi gases gives agreement with the
Pauli paramagnetism \cite{Lee}. In the magnetism of magnetized
pair-fermion gases, it is shown that the intrinsic spin play an
important role in relativistic paramagnetism or diamagnetism
\cite{Daicic}.

Besides the ideal gases, interaction need to be considered for
further understanding the magnetism of quantum gases. While the
Heisenberg model can usually be used to explain the magnetic
properties of quantum gases. Within the two different large-$N$
formulations, the low-temperature properties of quantum Heisenberg
models have been investigated \cite{Arovas}, both in ferromagnetic
(FM) and antiferromagnetic (AFM) situation. A Heisenberg's FM model
has also been used to deal with the high temperature susceptibility
\cite{Rushbrooke}.

Not only localized electrons, but also the ferromagnetism of
itinerant electrons has received a great deal of attention. The
development of itinerant electron magnetism has been outlined with
emphasis on spin fluctuations \cite{Moriya}. Research on itinerant
ferromagnetism of a trapped two-dimensional atomic gas has shown
that the effective interaction strength is unaffected by the
particle number density, although the FM phase is enhanced
\cite{Conduit2}. The FM phase transition of a two-dimensional
itinerant electrons Stoner Hamiltonian has been studied with quantum
Monte Carlo calculations. It is shown that a first-order FM
transition occurred for short screening lengths with a screened
Coulomb interaction \cite{Conduit1}. In spite of an infinitesimal
value of the coupling can induce a FM phase transition for spinor
Bose gases\cite{Gu}, the Stoner coupling of Fermi gases cannot lead
to a FM phase transition unless it is larger than a threshold. The
mechanisms of the Curie-Weiss law for the itinerant electron FM
material have been investigated through the $1/d$ expansion theory
\cite{Miyai}.

In this paper, by using the mean-field theory, the magnetic
properties of charged spin-$1/2$ Fermi gases with FM interactions
are investigated. Our results uncover a competition among
paramagnetism, diamagnetism and ferromagnetism. We also present a
comparison with the results of charged spin-1 Bose gas with FM
interactions which have been obtained previously \cite{Qin}. As the
increase of temperature, there is not a pseudo-critical temperature
for the charged spin-$1/2$ Fermi gases with FM interaction, which is
different from the case of Bose gas. The relationship of
susceptibility and temperature obey the Curie-Weiss law. In section
2, a model consisting of Landau diamagnetism, Pauli paramagnetism
and the FM effect is constructed. Then the magnetization density and
susceptibility are calculated respectively. Section 3 presents a
detailed discussion of obtained results. In section 4, we give a
brief summary.

\section{The Model}

We consider a spin-$1/2$ Fermi gas with FM couplings of N particles,
with the effective Hamiltonian written as
\begin{eqnarray}\label{Hamilt}
\bar{H}-\mu{N}=D_L\sum_{j,k_z,\sigma}\left(\epsilon^l_{jk_z}+\epsilon^{ze}_\sigma+\epsilon^m_\sigma-\mu\right){n}_{jk_z\sigma},
\end{eqnarray}
where $\mu$ is the chemical potential of the system. The quantized
Landau levels of the charged fermions are
\begin{eqnarray}\label{Deg}
\epsilon^l_{jk_z}=(j+\frac{1}{2})\hbar\omega+\frac{\hbar^2
k^2_z}{2m^\ast},
\end{eqnarray}
where $j=0,1,2,\ldots$ labels different Landau levels, and $\omega=q
B/(m^\ast c)$ is the gyromagnetic frequency, with charge $q$,
effective mass $m^{\ast}$ and the magnetic induction intensity $B$.
The degeneracy of the Landau levels is
\begin{eqnarray}\label{dege}
D_L=\frac{q B S}{2\pi\hbar c},
\end{eqnarray}
where $S$ is the section area of x-y plane of the system.

The Zeeman energy levels associated with the spin degree of freedom,
\begin{eqnarray}\label{Param}
\epsilon^{ze}_\sigma &=-g\hbar(q B/m^\ast c)\sigma
=-g\sigma\hbar\omega,
\end{eqnarray}
where $g$ is the Land\'{e} factor, and $\sigma$ denotes the spin-z index
of Zeeman state $\left| {F=1/2,m_F=\sigma} \right\rangle$ ($\sigma=1/2,
-1/2$).

The contribution to the effective Hamiltonian from the FM couplings
\cite{Tao} is
\begin{eqnarray}\label{Ferro}
\epsilon^{m}_\sigma=-4I\sigma(m+2\sigma n_{\sigma}),
\end{eqnarray}
where $I$ denotes FM coupling and spin polarization
$m=n_{\frac{1}{2}}-n_{-\frac{1}{2}}$.

Then we obtain the grand thermodynamic potential
\begin{align}\label{T1}
\Omega_{T\neq0}&=-\frac{1}{\beta}\ln\mathrm{Tr}e^{-\beta(\bar{H}-\mu{N})}
\nonumber \\
&=-\frac{1}{\beta}D_L\sum_{j,k_{z},\sigma}\ln\left[1+e^{-\beta\left(\epsilon^l_{jk_{z}}+\epsilon^{ze}_\sigma+\epsilon^{m}_\sigma-\mu\right)}\right],
\end{align}
where $\beta=1/(k_B T)$. Through the Taylor expansions and integral
over $k_z$, we have
\multlinegap=25pt
\begin{multline}\tag{7}
\Omega_{T\neq0}=-\frac{\omega
V}{\hbar^{2}}(\frac{m^\ast}{2\pi\beta})^\frac{3}{2}\sum^{\infty}_{l=1}\\\shoveleft{\sum_{\sigma}\frac{(-1)^{l+1}l^{-\frac{3}{2}}
e^{-l\beta\left[\frac{1}{2}\hbar\omega-g\sigma\hbar\omega-4I\sigma(m+2\sigma
n_\sigma)-\mu\right]}}{1-e^{-l\beta\hbar\omega}}},
\end{multline}
where $V$ is the volume of the system. A similar treatment has been
used to deal with the diamagnetism of scalar Bose gases
\cite{Toms1,Toms2,Standen}. We have extended it to further study the
competition between diamagnetism and paramagnetism of charged spin
quantum gases \cite{Jian1,Jian2}. As far as FM interaction is
concerned, it is shown that the mean-field theory is still effective
in understanding the main physics of magnetism \cite{Gu,Qin,Tao}.

Some compact notations for the class of sums are introduced for simplicity,
\begin{align}\tag{8}
F^\sigma_\tau[\alpha,\delta]=\sum^{\infty}_{l=1}\frac{(-1)^{l+1}l^{\alpha/2}e^{-l\beta\hbar\omega\left[\frac{1}{2}-g\sigma-\frac{4I
\sigma(m+2\sigma
n_\sigma)}{\hbar\omega}-\frac{\mu}{\hbar\omega}+\delta\right]}}{{\left(1-e^{-l\beta\hbar\omega}\right)}^\tau}.
\end{align}

Then we may rewrite equation (7) as
\begin{align}\tag{9}
\Omega_{T\neq0}=-\frac{\omega
V}{\hbar^2}(\frac{m^\ast}{2\pi\beta})^\frac{3}{2}\sum_{\sigma}F^\sigma_1[-D,0],
\end{align}
where $D=3$ is the space dimensionality.

Then the particle number density $n=N/V$ can be obtained through the grand thermodynamic potential,
\multlinegap=80pt
\begin{multline}\tag{10}
n=-\frac{1}{V}\left(\frac{\partial\Omega_{T\neq0}}{\partial{\mu}}\right)_{T,V}
\\\shoveleft{=x(\frac{m^\ast}{2\pi\beta\hbar^2})^\frac{3}{2}\sum_{\sigma}F^\sigma_1[-1,0]},
\end{multline}
where $x=\beta\hbar\omega$.

Taking the grand thermodynamic potential derivative with respect to the magnetic induction intensity $B$, the total magnetization density can be obtained
\multlinegap=0pt
\begin{multline}\tag{11}
M_{T\neq0}=-\frac{1}{V}\left(\frac{\partial\Omega_{T\neq0}}{\partial{B}}\right)_{T,V}
\\\shoveleft{=\frac{q\hbar}{m^\ast c}(\frac{m^\ast}{2\pi\beta\hbar^2})^\frac{3}{2}}\\\sum_{\sigma}
\left\{F^\sigma_1[-3,0]+x\left[\left(g\sigma-\frac{1}{2}\right)F^\sigma_1[-1,0]-F^\sigma_2[-1,1]\right]\right\}.
\end{multline}

In external magnetic field $H$, we have
\begin{align}\tag{12}
B=H+4\pi M.
\end{align}

It is convenient to introduce some dimensionless parameters, such as $t=T/{T^\ast}, \bar M={m^\ast c M}/(n \hbar q), \bar\omega={\hbar\omega}/(k_BT^\ast), \bar I={In}/(k_BT^\ast), \bar\mu={\mu}/(k_BT^\ast), \bar m={m}/{n}, \bar n_\sigma={n_\sigma}/{n}, h={q \hbar H}/(m^\ast c k_BT^\ast)$,
and $x=\bar\omega/t$, the characteristic temperature of the system $T^\ast$ is given by $k_BT^\ast={2\pi\hbar^2 n^\frac{2}{3}}/{m^\ast}$. Accordingly, we can re-express equations (10), (11) and (12),

\begin{align}\tag{13a}
1=\bar\omega t^\frac{1}{2}\sum_\sigma\bar F^\sigma_1[-1,0],
\end{align}
\multlinegap=0pt
\begin{multline}\tag{13b}
\bar M_{T\neq0}=t^\frac{3}{2}\\\sum_\sigma \left\{ \bar
F^\sigma_1[-3,0] +\frac{\bar\omega}{t} \left[
\left(g\sigma-\frac{1}{2}\right)\bar F^\sigma_1[-1,0]-\bar
F^\sigma_2[-1,1] \right] \right\},
\end{multline}
\begin{align}\tag{13c}
\bar\omega=h+4\pi\gamma\bar M,
\end{align}
where $\gamma=(q^2 n^{1/3})/(2\pi m^\ast c^2)$, and

\begin{align}\tag{14}
\bar
F^\sigma_\tau[\alpha,\delta]=\sum^{\infty}_{l=1}\frac{(-1)^{l+1}l^{\alpha/2}e^{-l\frac{\bar\omega}{t}\left[\frac{1}{2}
-g\sigma-\frac{4\bar I\sigma(\bar m+2\sigma\bar
n_\sigma)}{\bar\omega}-\frac{\bar\mu}{\bar\omega}+\delta\right]}}
{{\left(1-e^{-l\frac{\bar\omega}{t}}\right)}^\tau},
\end{align}
where $\bar\mu$ is the dimensionless parameter of the chemical potential, which can be determined from the mean-field
self-consistent calculations.

At last, we calculated the susceptibility of charged Fermi gases with FM couplings.
From the formula $\displaystyle\chi_{M}=(\frac{\partial{M}}{\partial{H}})_{T,V}$, the derivation of equation (11) for the magnetic field $H$,
the expression of susceptibility can be obtained as follows,
\begin{align}\tag{15}
\chi_{M}=\frac{\gamma b}{\bar\omega-4\pi\gamma b},
\end{align}
with
\multlinegap=5pt
\begin{multline}\tag{16}
b=2\bar\omega t^{1/2}\sum_\sigma\{(g\sigma-\frac{1}{2})\bar
F^\sigma_1[-1,0]-\bar
F^\sigma_2[-1,1]\}-\bar\omega^2t^{-1/2}\\\shoveleft{\sum_\sigma\{2(g\sigma-1)\bar
F^\sigma_2[1,1]-(g\sigma-\frac{1}{2})^2\bar F^\sigma_1[1,0]-2\bar
F^\sigma_3[1,2]\}}.
\end{multline}

\section{Results and discussions}

\begin {figure}[t]
\center\includegraphics[width=0.45\textwidth,keepaspectratio=true]{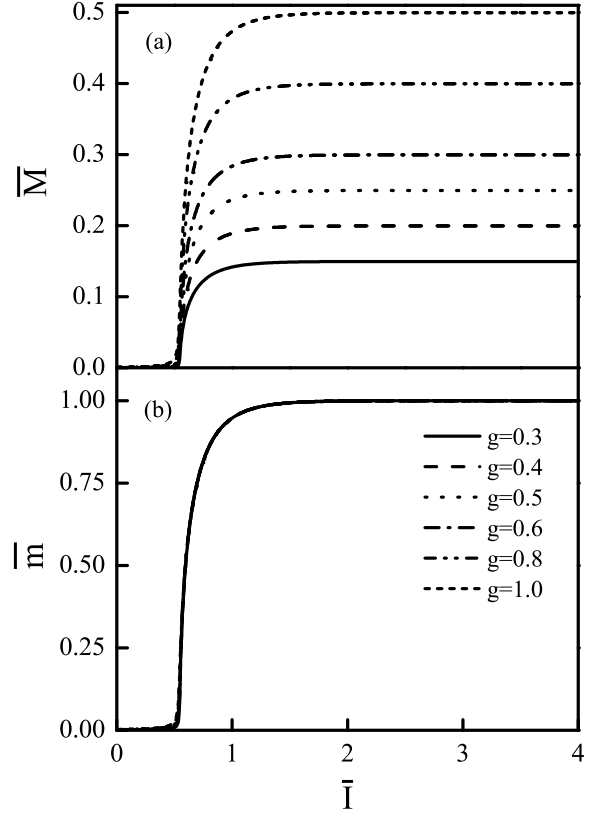}
\caption{(a) The reduced magnetization density $\bar M$, (b) $\bar m=\bar n_{\frac{1}{2}}-\bar n_{-\frac{1}{2}}$ as a function of reduced FM
coupling constant $\bar I$ at reduced temperature $t=1.5$ and the reduced
magnetic field $h=0.005$. The Land\'{e} factor $g$ is chosen as 0.3 (solid line), 0.4 (dashed line), 0.5 (dotted line),
0.6 (dash dotted line), 0.8 (dash dot dotted line), and 1.0 (short dashed line). \label{fig1}}
\end{figure}

In the following discussions we will focus on the competition of
magnetism, and explore the factors that determine the magnetic
competition. Meanwhile a comparison with the results of Bose gases
with FM coupling will also be presented.

Firstly, the dimensionless magnetization density $\bar M$ and $\bar
m=\bar n_{1/2}-\bar n_{-1/2}$ versus $\bar I$ is shown in figure 1.
$\bar I_c$ is used to describe the critical value of reduced FM
coupling constant of the paramagnetic phase to ferromagnetic phase
transition. We can find the crossover of $\bar M$ from figure 1(a).
When the reduced FM coupling constant $\bar I$ is smaller than $\bar
I_c$, $\bar M$ is always equal to zero. And then $\bar M$ begins to
increase with increasing $\bar I$ from $\bar I_c$. It suggests that
there exists a spontaneous magnetization with the increase of $\bar
I$ in the weak magnetic field. Figure 1(b) indicates that although
the value of land\'{e} factor $g$ changes, the evolvement of $\bar
m$ versus $\bar I$ almost overlaps. And the value of $\bar I_c$ is
identical for the given reduced temperature, in despite of the
values of land\'{e} factor are different. While the critical value
of FM coupling constant $\bar I_c$ approximates to 0.5, which can be
evaluated from figure 1. The internal field comes from the
spontaneous magnetization results in the diamagnetism \cite{Qin}.
Figure 1 shows that ferromagnetism exceeds the diamagnetism with
increasing $\bar I$.

Since $\bar I_c$ is an important parameter in the transformation
between FM phase and paramagnetic phase. The threshold of $\bar
I_{c}$ vs temperature is plotted in figure 2. The FM phase situates
above $\bar I_c$, while the paramagnetic phase locates under the
$\bar I_c$. We can find that the value of $\bar I_c$ increases
linearly with the increase of temperature. The results is similar to
the Bose gas with FM coupling \cite{Qin}, although they submit to
different statistical rules, respectively. It indicates that the
crossover from paramagnetic phase to FM phase is more difficult with
increasing the temperature.

\begin {figure}[t]
\center\includegraphics[width=0.45\textwidth,keepaspectratio=true]{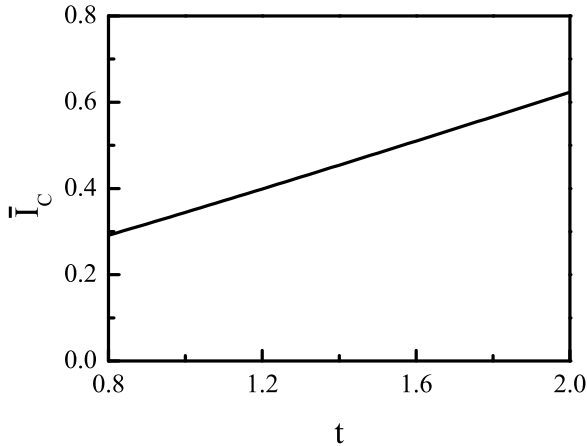}
\caption{$\bar I_c$ versus reduced temperature t phase diagram of
charged spin-$1/2$ Fermi gases at magnetic field $h=0.005$. \label{fig2}}
\end{figure}

\begin {figure}[t]
\center\includegraphics[width=0.45\textwidth,keepaspectratio=true]{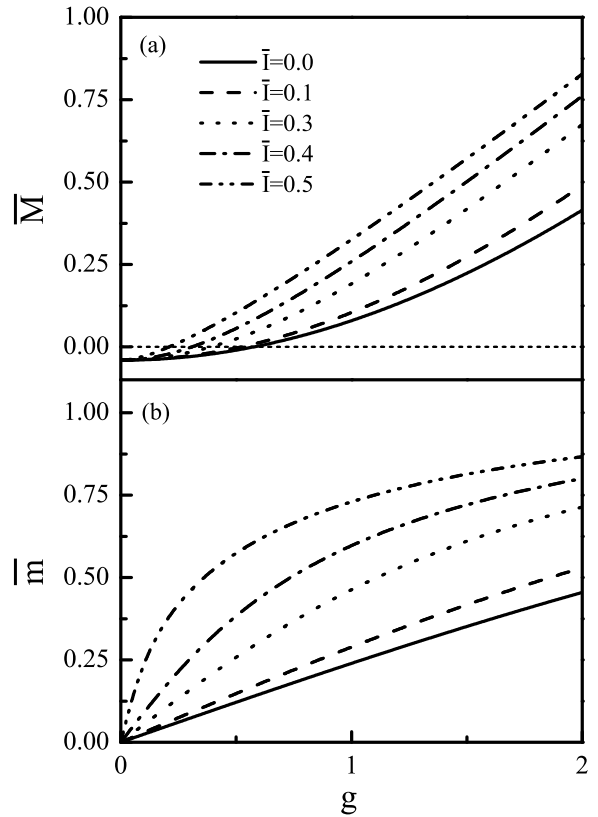}
\caption{(a) The reduced magnetization density $\overline{M}$, (b) $\bar m=\bar n_{1/2}-\bar n_{-1/2}$ versus Land\'{e} factor $g$ at
reduced temperature $t=1.5$ and $h=0.8$, where $\bar I=$0.0 (solid line), 0.1 (dashed line), 0.3 (dotted line),
0.4 (dash dotted line), and 0.5 (dash dot doted line).\label{fig3}}
\end{figure}

\begin {figure}[t]
\center\includegraphics[width=0.45\textwidth,keepaspectratio=true]{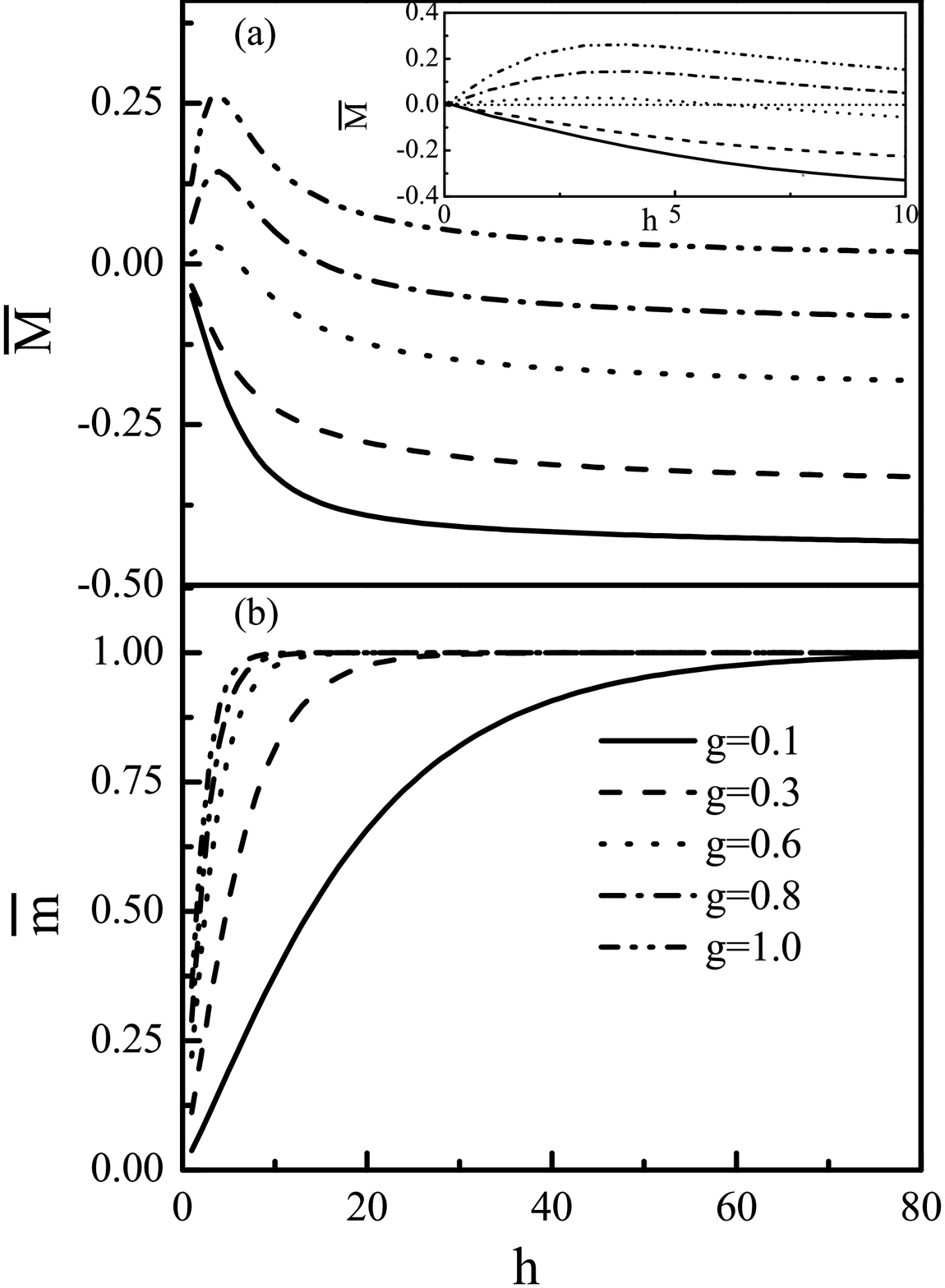}
\caption{(a) The reduced magnetization density $\overline{M}$, (b) $\bar m=\bar n_{1/2}-\bar n_{-1/2}$ as a function of reduced magnetic field $h$ at $t=1.5$ and $\overline{I}=0.1$, where Land\'{e} factor $g$=0.1(solid line), 0.3 (dashed line), 0.6 (dotted line), 0.8 (dash dotted line), and 1.0 (dash dot dotted line). The inset of figure 4(a) depicts the relationship of reduced magnetization density $\bar M$ versus magnetic field $h$ with the magnetic field region lies between 0 and 10. \label{fig4}}
\end{figure}

After studying the spontaneous magnetization in the weak magnetic
field, figure 3 is plotted in order to understand the influence of
Land\'{e} factor $g$. In figure 3, the magnetic field is chosen as
$h=0.8$ and the reduced temperature $t=1.5$. The horizontal line
$\bar M=0$ in figure 3(a) is used to distinguish the region of
paramagnetism and diamagnetism. It is shown that $\bar M$ is a
negative value when $g$ is small, which mainly derives from the
diamagnetic contribution. From figure 3(b) we can find that $\bar m$
increase with increasing of Land\'{e} factor $g$ until approximates
saturation at higher $\bar I$. It indicates that the larger $\bar I$
the larger $\bar m$ for the identical value of Land\'{e} factor $g$,
where the Land\'{e} factor denotes the intensity of the
paramagnetism \cite{Jian1,Jian2}. From figure 3(a) and 3(b), we can
find that the enhance of ferromagnetism stimulates the increasing of
paramagnetism. That is ferromagnetism cooperates with paramagnetism
to confront diamagnetism in a finite magnetic field. This is
different from the result of the Bose gas with FM coupling
\cite{Qin}. Where it is faintly affected by the FM coupling in the
evolvement of magnetization density with $g$.

\begin {figure}[t]
\center\includegraphics[width=0.45\textwidth,keepaspectratio=true]{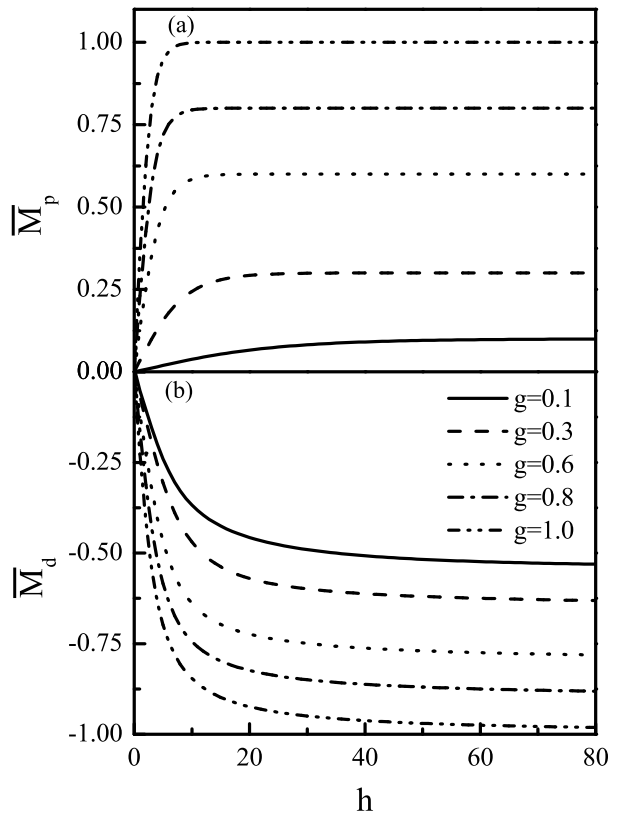}
\caption{(a) The reduced paramagnetization density $\overline{M}_p$ , (b) the reduced diamagnetization density $\bar M_d$ as
a function of reduced magnetic field $h$ at $t=1.5$ and $\overline{I}=0.1$, where Land\'{e} factor $g$=0.1 (solid
line), 0.3 (dashed line), 0.6 (dotted line), 0.8 (dash dotted line), and 1.0 (dash dot dotted line). \label{fig5}}
\end{figure}

Figure 4 is plotted to get more insight of the dependence of the
magnetization density on magnetic field. The system presents
diamagnetism when $g$ is small. Furthermore, there exists a
threshold $g_c$ when $0.3<g<0.6$, which making $\bar M<0$ while
$g<g_c$. This is independent of the magnetic field. From the inset
of figure 4(a), we can find that the magnetization density
approximates to a small positive value at first when $g=0.6$. While
$g\geq0.6$, the magnetization density increases firstly with
increasing the magnetic field, and then declines up to reach a
saturation value. It is shown that the paramagnetism competes with
diamagnetism in this definite FM coupling situation. The increasing
of Land\'{e} factor promotes paramagnetism, while the magnetic field
facilitates diamagnetism.

\begin {figure}[t]
\center\includegraphics[width=0.45\textwidth,keepaspectratio=true]{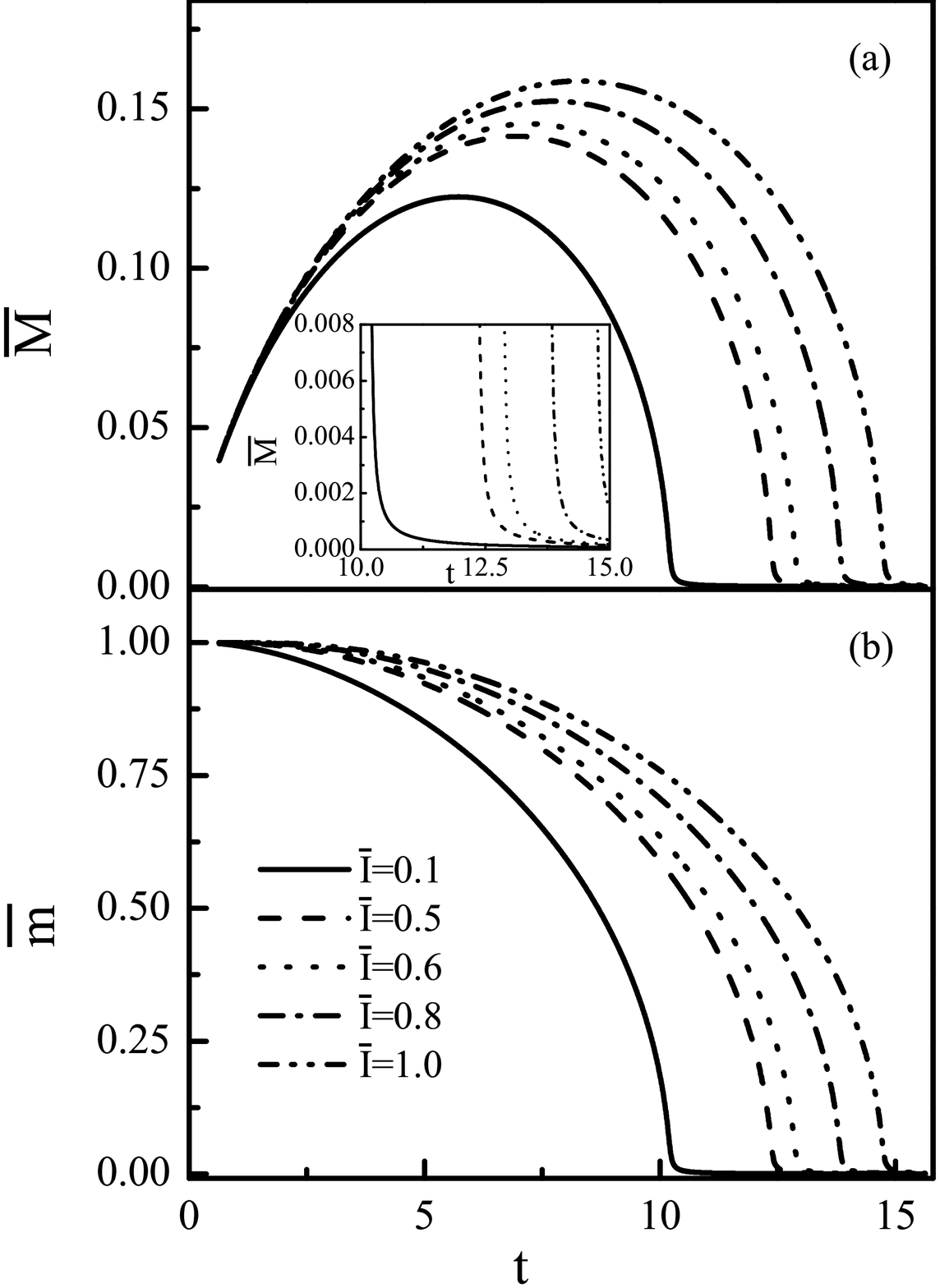}
\caption{(a) The reduced magnetization density $\bar M$, (b) $\bar m=\bar n_{1/2}-\bar n_{-1/2}$ versus reduced
temperature $t$ at reduced magnetic field $h=0.005$ and Land\'{e} factor $g=0.8$, where $\overline{I}$=0.1 (solid line), 0.5 (dashed line), 0.6 (dotted line), 0.8 (dash dotted line), and 1.0 (dash dot dotted line). The inset of figure 6(a) depicts the smooth decline of reduced magnetization
density $\bar M$ with the reduced temperature lies between 10 and 15. \label{fig6}}
\end{figure}

\begin {figure}[t]
\center\includegraphics[width=0.45\textwidth,keepaspectratio=true]{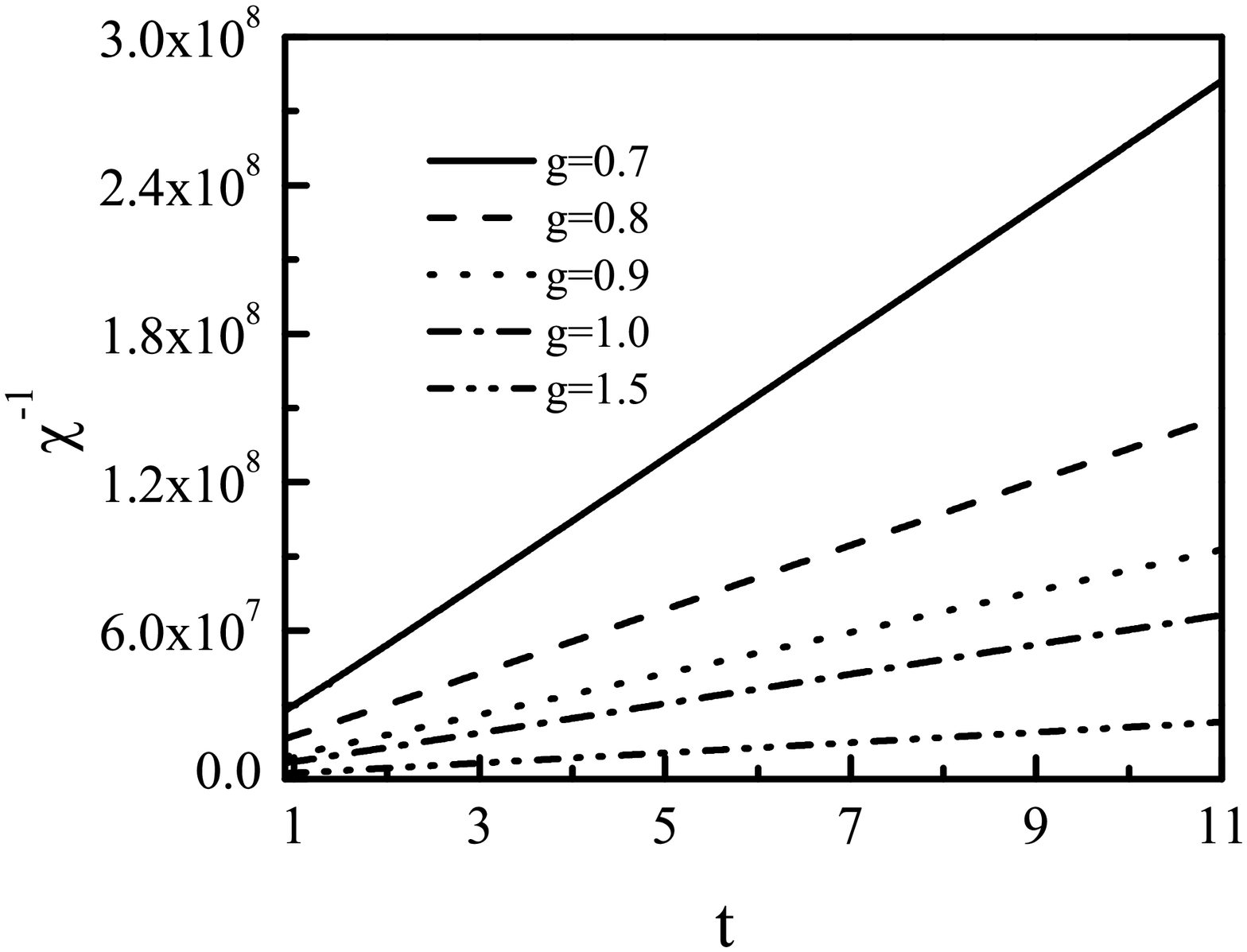}
\caption{ The reciprocal of susceptibility $1/\chi$ versus reduced temperature $t$ at reduced magnetic field $h=0.005$ and reduced FM coupling constant $\bar I$=0.5. The Land\'{e} factors: g=0.7 (solid line), 0.8 (dashed line), 0.9 (dotted line), 1.0 (dash dotted line), and 1.5 (dash dot dotted line). \label{fig7}}
\end{figure}

For a more detailed understanding of the paramagnetism and
diamagnetism respectively, now we turn to examine the
paramagnetization density $\bar M_p=g \bar m$ and diamagnetization
density $\bar M_d=\bar M-\bar M_p$ in figure 5. It is shown that
both the paramagnetization density and diamagnetization density
increases with enhancing the magnetic field at first. And then the
paramagnetization density $\bar M_p$ tends to saturate when the
magnetic field intensity is strong. While the absolute value of
diamagnetization density $|\bar M_d|$ increases with the increase of
magnetic field intensity. When the magnetic field continues to
increase, the diamagnetism is close to a saturated value. Therefore
for the higher magnetic field the total magnetization density $\bar
M$ reaches to saturated values. This is in conformity with the
result of figure 4 in qualitatively. With the increase of magnetic
field, the contribution come from ferromagnetism, paramagnetism and
diamagnetism reach to maximum, so the peak of total magnetization
density appears when Land\'{e} factor $g$ is 0.8 and 1.0 in figure
4. Whereas the diamagnetism increases faster than paramagnetism as
the magnetic field continues to increase. It accounts for the total
magnetization density decreases with the increase of magnetic field
until nearly saturated, which can be found in figure 4.

The characteristic parameter $\gamma$ has been set to $10^{-6}$ in
figures 1$\thicksim$5, where the particle number density is
$8/nm^{3}$ and the charge and mass are evaluated from a thin
electron gas. To further investigate the FM phase transition of the
charged spin-$1/2$ Fermi gases in a broad temperature region
including low temperature, $\gamma=10$ is assumed. Figure 6 shows
$\bar M$ and $\bar m$ as a function of the temperature $t$ when
$g=0.8$ in magnetic field $h=0.005$. It denotes that the maximal
$\bar M$ occurs at a definite temperature, and then decreases in
both lower temperature and higher temperature regimes. Moreover,
$\bar M$ decreases faster in the low temperature region than the
case of the Bose gas with FM coupling \cite{Qin}. It reflects the
diamagnetism strengthens greatly in this region for Fermi gas. With
increasing the temperature, a flat decline appears when $\bar M$ is
close to zero, which can be seen clearly from the inset of figure
6(a). This is obviously different from the Bose gas. In our previous
study on Bose gas with FM coupling \cite{Qin}, a sharp decline
emerges when $\bar M$ approaches to zero, which suggests that there
exists a pseudo-condensate temperature in the transition from
ferromagnetism to paramagnetism. However, at high temperature region
for the charged Fermi gases with FM interaction, $\bar M$ and $\bar
m$ are asymptotic with respect to the zero point for different
values of $\bar I$. This demonstrates that there does not exist the
pseudo-critical temperature for Fermi gases. The difference between
Fermi gases and Bose gases may be attributed to the different
statistical distribution.

Figure 7 plots the evolution of reciprocal of susceptibility with
the reduced temperature $t$, where the characteristic parameter
$\gamma$ is reset as $10^{-6}$. It is shown that the curves of
$1/\chi$ is proportional to the reduced temperature $t$. This
suggests that the susceptibility of Fermi gases with FM coupling
conforms to the Curie-Weiss law. When the temperature is fixed, the
susceptibility increases with the increase of the Land\'{e} factor
$g$. Since the Land\'{e} factor $g$ denotes the strength of
paramagnetism, it suggests that the increase of susceptibility is
mainly attributed to the contribution of paramagnetism at the fixed
temperature and magnetic field.

\section{Summary}
This paper has revealed the interplay among paramagnetism,
diamagnetism and ferromagnetism of the charged spin-$1/2$ Fermi
gases with FM interaction. The paramagnetic effect is described by
the Land\'{e} factor $g$. Our results show that the ferromagnetism
overcomes the diamagnetism when $\bar I>\bar I_{c}$ in weak magnetic
field. $\bar I_c$ increases linearly with the increase of
temperature. With increasing $g$, the gas presents paramagnetism
from diamagnetism. The reduced magnetization density $\bar M$
declines in the high magnetic field region, which indicates that the
contribution comes from diamagnetism enhances in the region of
strong magnetic field. As the increase of temperature, the
magnetization density asymptotically approximates to zero, which is
different from the case of Bose gas with FM coupling. It indicates
that there is not a pseudo-critical temperature for charged Fermi
gases with FM interaction. The diagram of susceptibility is in
accordance with the Curie-Weiss law.

\begin{acknowledgement}
We would like to thank Professor Qiang Gu for the helpful
discussions. BW and JQ are supported by the Fundamental Research
Funds for the Central Universities under Grant No. FRF-TP-14-074A2,
the Beijing Higher Education Young Elite Teacher Project under Grant
No. 0389, and the National Natural Science Foundation of China under
Grant No. 11004006, and HG is supported by the National Natural
Science Foundation of China under Grant No. 11274032.
\end{acknowledgement}

All authors contributed equally to this paper.

\end{document}